\documentclass[12pt]{article}
\usepackage{times}
\usepackage{geometry}
\usepackage[utf8]{inputenc}
\usepackage{enumitem,amssymb}
\usepackage{ragged2e}
\usepackage{natbib}
\usepackage{graphicx}
\usepackage{color}
\usepackage{amsmath,amssymb}
\newlist{thematic}{itemize}{8}
\setlist[thematic]{label=$\square$}
\usepackage{pifont}
\newcommand{\cmark}{\ding{51}}%
\newcommand{\done}{\rlap{$\square$}{\raisebox{2pt}{\large\hspace{1pt}\cmark}}%
\hspace{-2.5pt}}

\newcommand{\degr}{^{\circ}}

\begin{document}
\raggedright
\huge
Astro2020 Science White Paper \linebreak

Measurement of the Free-Floating Planet Mass Function with Simultaneous {\it Euclid} and {\it WFIRST} Microlensing Parallax Observations\linebreak
\normalsize

\noindent \textbf{Thematic Areas:} \hspace*{60pt} $\done$ Planetary Systems \hspace*{10pt} $\square$ Star and Planet Formation \hspace*{20pt}\linebreak
$\square$ Formation and Evolution of Compact Objects \hspace*{31pt} $\square$ Cosmology and Fundamental Physics \linebreak
  $\square$  Stars and Stellar Evolution \hspace*{1pt} $\square$ Resolved Stellar Populations and their Environments \hspace*{40pt} \linebreak
  $\square$    Galaxy Evolution   \hspace*{45pt} $\square$             Multi-Messenger Astronomy and Astrophysics \hspace*{65pt} \linebreak
  
\textbf{Principal Author:}

Name: Matthew T. Penny
 \linebreak						
Institution: The Ohio State University
 \linebreak
Email: penny@astronomy.ohio-state.edu
 \linebreak
Phone: 614-292-6925
 \linebreak
 
 \textbf{Co-authors:}

 Etienne Bachelet (Las Cumbres Observatory),
 Samson Johnson (The Ohio State University),
 Jean-Phillipe Beaulieu (University of Tasmania \& Institut d'Astrophysique de Paris),
 Eamonn Kerins (University of Manchester),
 Jason Rhodes (Jet Propulsion Laboratory),
 Rachel Akeson (Caltech-IPAC),
 David Bennett (NASA Goddard Space Flight Center \& The University of Maryland),
 Charles Beichman (IPAC/NExScI),
 Aparna Bhattacharya (University of Maryland College Park \&
NASA Goddard Space Flight Center),
Valerio Bozza (Universit\`a di Salerno),
 Sebastiano Calchi Novati (IPAC/Caltech),
 B. Scott Gaudi (The Ohio State University),
 Calen B. Hendederson (Caltech/IPAC),
Shude Mao (Tsinghua University),
 Radek Poleski (The Ohio State University),
 Cl\'ement Ranc (NASA Goddard Space Flight Center),
 Kailash C. Sahu (STScI),
 Yossi Shvartzvald (IPAC),
 Rachel Street (Las Cumbres Observatory)
 \linebreak

 \newcommand{\euclid}{{\it Euclid}}
 \newcommand{\wfirst}{{\it WFIRST}}
 \newcommand{\mearth}{M_{\oplus}}
 \newcommand{\msun}{M_{\odot}}
 \newcommand{\reproj}{\tilde{r}_{\mathrm{E}}}
 \newcommand{\thetae}{\theta_{\mathrm{E}}}
   \newcommand{\pie}{\pi_{\mathrm{E}}}

   {\bf Abstract  (optional):}

   Free-floating planets are the remnants of violent dynamical rearrangements of planetary systems. It is possible that even our own solar system ejected a large planet early in its evolution. \wfirst\ will have the ability to detect free-floating planets over a wide range of masses, but it will not be able to directly measure their masses. Microlensing parallax observations can be used to measure the masses of isolated objects, including free-floating planets, by observing their microlensing events from two locations. The intra-L2 separation between \wfirst\ and \euclid\ is large enough to enable microlensing parallax measurements, especially given the exquisite photometric precision that both spacecraft are capable of over wide fields. In this white paper we describe how a modest investment of observing time could yield hundreds of parallax measurements for \wfirst's bound and free-floating planets. We also describe how a short observing campaign of precursor observations by \euclid\ can improve \wfirst's bound planet and host star mass measurements.

\pagebreak

\justify
{\bf Free-floating planets constrain planetary system evolution}

A full understanding of planet formation and planetary system evolution must account for not only those systems that are created and survive, but also those that are destroyed.

   At the end of the planet formation process, as the protoplanetary disk is photoevaporated away, chaos is unleashed within the newly formed planetary systems. The ejection of planets from their systems, or into much larger orbits, is a natural consequence of the $N$-body dynamics of newly formed planetary systems~\citep[e.g.,][]{Rasio1996,Weidenschilling1996}, whose mutual interactions had previously been damped by gas and dust disks. Extensions of the Nice model of the early evolution of our solar system~\citep{Gomes2005,Tsiganis2005,Morbidelli2005}, hypothesize that initially five giant planets were formed before one Neptune-like planet was ejected from the solar system after a close encounter with Jupiter~\citep{Nesvorny2011,Batygin2012}. Even if not completely ejected, a large planet could still lurk undetected in the outer reaches of the solar system~\citep[e.g.,][]{Trujillo2014,Batygin2016,Batygin2019}. In addition to large planets, vast quantities of planetesimals of various sizes are likely to be ejected from planetary systems, possibly up to tens of Earth masses in total, perhaps with up to an Earth mass in large Pluto-sized objects~\citep[e.g.,][]{Levison2008}. Theoretical expectations for the ejection rate of small and large planets vary~\citep[e.g.,][]{Mann2010,Pfyffer2015,Barclay2017}, but recently a small number of strong free-floating planet candidates have been detected by microlensing surveys~\citep{Mroz2018, Mroz2019}.

   A measurement of the mass function of free-floating and widely bound planets therefore places important constraints on the process of planet formation and evolution, and the frequency of violent instabilities that lead to planetary ejections.\\

   \noindent
   {\bf Free-floating planet mass measurements from {\it Euclid}-{\it WFIRST} L2-L2 parallax}
   
   \wfirst\ will detect free-floating and wide-orbit planets beyond 10~AU over a wide range of masses. Should ${\sim}1\mearth$-worth of Pluto-mass objects, or ${\sim}0.1\mearth$-worth of $0.01$--$1\mearth$ objects, exist per star on wide orbits or ejected from their systems, \wfirst's microlensing survey will detect them~\citep[][Johnson et al. 2019 in prep.]{Spergel2015}. If they exist with a mass function similar to microlensing's bound planets, \wfirst\ will detect hundreds. However, \wfirst\ alone will provide only order-of-magnitude constraints on the mass of each object, based on their microlensing timescale, which is a degenerate combination of the object's mass, distance, and velocity relative to the microlensing source star. For some fraction of \wfirst's free-floating planets, a further constraint on mass and distance to the object is provided by finite source effects, which yield the angular Einstein ring radius $\thetae$ if the color of the microlensing source star is also measured. But $\thetae$ it is not sufficient to completely resolve the mass degeneracy.

   A complete solution to the mass and distance of a free floating planet can be made if, in addition to $\thetae$, the microlensing parallax $\pie$ is also measured~\citep[e.g.][]{Gould2000}. For a free-floating planet, which causes a very short $\lesssim1$~day, microlensing event, the only way to measure $\pie$ is by simultaneous high-cadence observation of the event by two observers separated by a baseline that is smaller than, but comparable to the Einstein radius projected onto the solar system, $\reproj$. For typical values of microlensing parameters, $\reproj$ takes values
   \begin{equation}
     \reproj \sim 10 \textrm{AU} \sqrt{M/\msun} \sim 0.03 \textrm{AU} \sqrt{M/\mearth}.
   \end{equation}
   This means that the ${\sim}0.01$~AU baseline between \wfirst\ at L2 and the Earth is well suited to measuring microlensing parallax of free-floating planets. Combining $\pie$ and $\thetae$ measurements yields a direct measurement of the lens' mass
\begin{equation}
  M = \frac{\thetae}{\kappa\pie},
\end{equation}
where $\kappa = 8.144$~mas$/\msun$ is a constant. 

   Unfortunately however, the majority of \wfirst's microlensing events will be too faint to be seen from ground-based telescopes due to their worse resolution and higher backgrounds~\citep{Penny2019}, and parallax measurements will probably only be possible for the brightest events, even with telescopes such as LSST~\citep[e.g.][]{Street2018}.

\begin{figure}
  \includegraphics[width=0.5\textwidth]{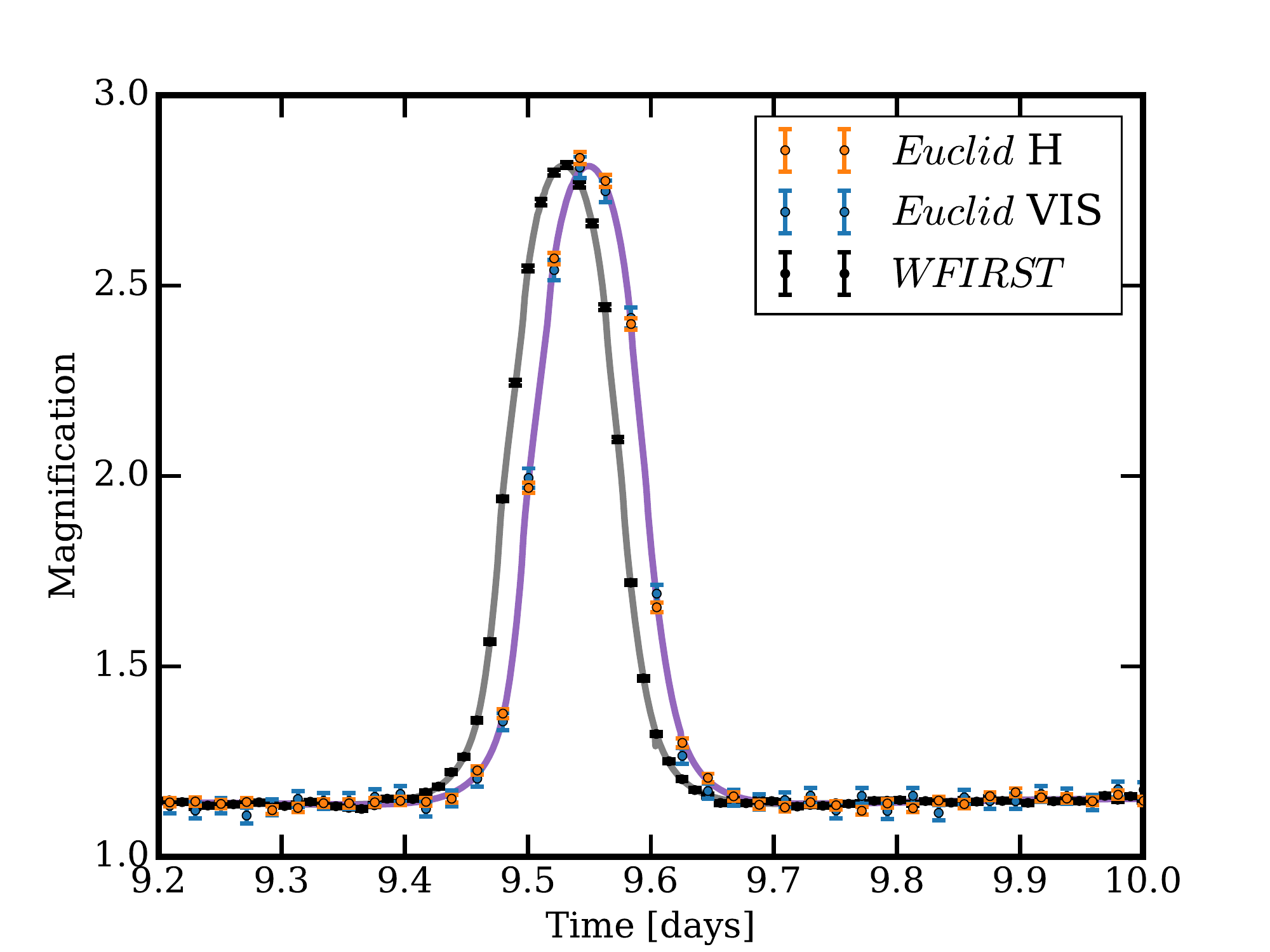}
  \includegraphics[width=0.5\textwidth]{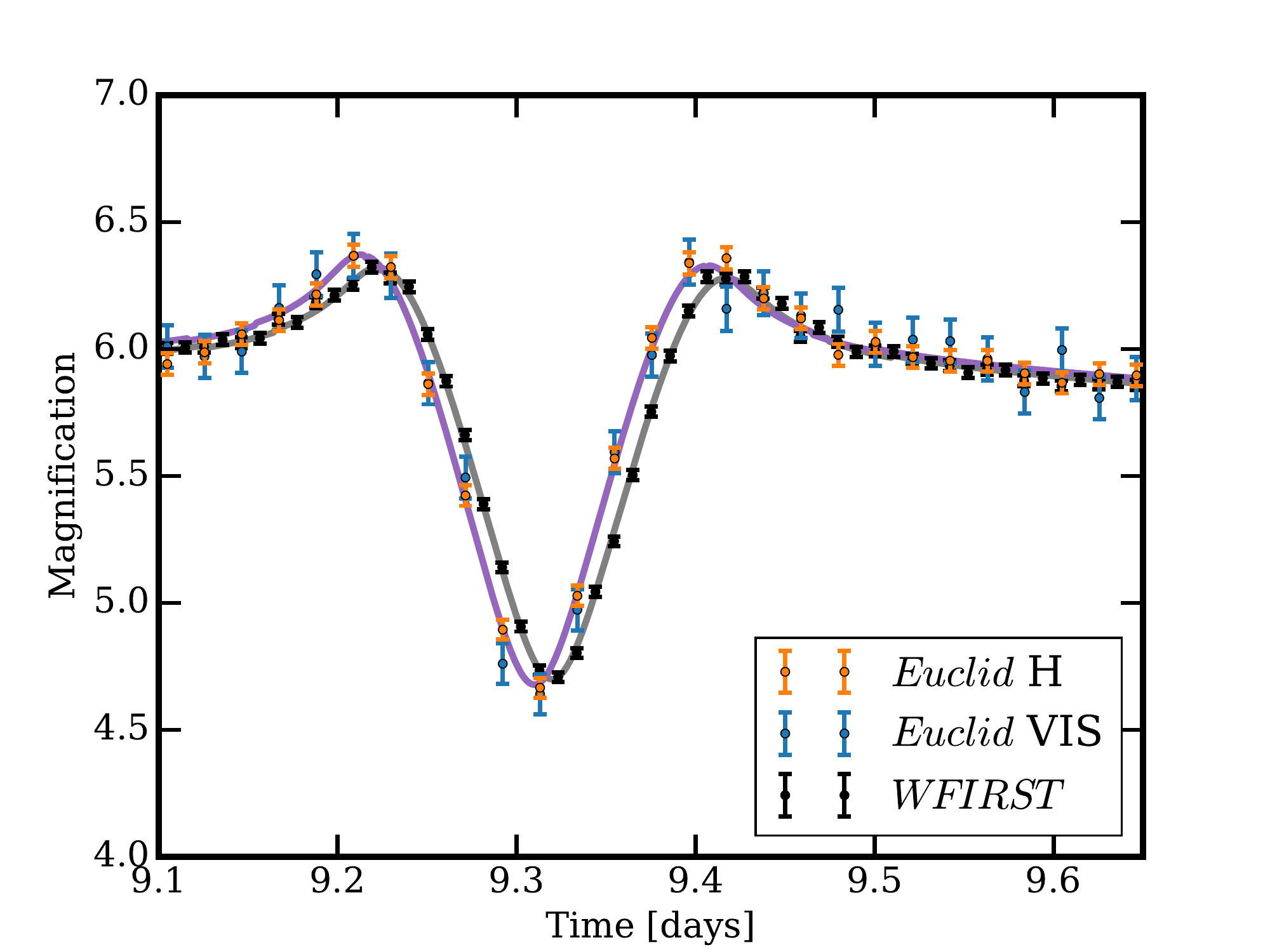}
  \caption{Simulated microlensing lightcurves demonstrating how even the short intra-L2 baseline between \wfirst\ and \euclid\ can produce measurable differences between the lightcurves of Earth-mass planetary microlensing events. The left plot shows the lightcurve of a $1\mearth$ wide-orbit planet with a lightcurve resembling a free-floating planet, whereas the right plot shows the lightcurve of a $1\mearth$ planet 1~AU from its $0.11\msun$ host.}
\end{figure}
   
   Thankfully, the orbits taken by spacecraft around L2 are large, and can be a significant fraction of the Earth-L2 distance itself. The only other telescope operating at L2 and capable of high-cadence observations over a wide field of view will be \euclid~\citep{Laureijs2011}, which \citet{Penny2013} demonstrated would be highly capable of conducting its own microlensing survey. \euclid's orbit about L2 will have a diameter of the order of $1$~million km, or ${\sim}0.007$~AU. Figure~1 shows a simulated lightcurve of an Earth-mass free-floating planet observed simultaneously from \wfirst\ and \euclid\ at two different locations within L2. Microlensing parallax has a large impact on how the microlensing event appears to both spacecraft. \euclid's high-cadence observations in two colors would also guarantee the measurement of the source star color, which is required to measure $\thetae$; \wfirst\ may not be able to measure source colors for all free-floating planets because it is expected to gather color measurements at a lower cadence, \citep[${\sim}6$--$12$~hours]{Penny2019}.\\

   \noindent
{\bf \euclid\ enables the measurement of parallaxes and masses for bound planet and their hosts}

\wfirst\ will measure host and planet masses in most of its microlensing events via direct detection of light from the lens star as it separates from the microlensing source star~\citep[][see also the white paper by Bhattacharya et al.]{Spergel2015,Bennett2007,Bhattacharya2018}. Lens detection provides a measurement of the relative proper motion of the lens and source, which when combined with the event timescale and an assumption of the source distance, yields a mass-distance relation (mass as a function of unknown lens distance). Lens detection also allows an estimate of the lens magnitude, which with the assumption of its extinction, and that there are no other significant contributors to light that affect the measurement (e.g., a luminous companion to the lens or source, or an unrelated star), yields another mass-distance relation. These can be combined to produce a unique solution for the lens mass and distance, even for isolated stars. However, both mass-distance relations can have similar functional forms for certain combinations of observables, which can lead to large uncertainties on the lens mass and distance in some cases. Planetary events with measurable finite source effects provide an independent estimate of $\theta_{\rm E}$ and relative lens-source proper motion, which can mitigate the effect of excess light from unknown companions, and thus more secure and accurate mass and distance estimates. Even with this, however, direct detection of the lens can fail if the relative proper motion is too low, or the source too bright, to allow detection of the lens.

Therefore, it is important to have an independent method with which to estimate or constrain the lens mass. Measurement of microlensing parallax $\pi_{\rm E}$ provides another mass-distance relation that varies inversely with distance, i.e., in the opposite sense to the previously mentioned mass distance relations, which can yield a more precise mass measurement. Additionally, if finite source effects are measured, the lens mass can be written purely in terms of observables, $M=\theta_{\rm E}/\kappa\pi_{\rm E}$, that require no astrophysical assumptions (e.g., a mass-luminosity relation, source distance, etc.).

The same scaling arguments for the scale of the projected Einstein radius apply to bound-planets as well as free-floating planets. Therefore, {\bf high-cadence observations from \euclid\ can also be used to measure microlensing parallax from the planetary microlensing feature of bound planets}, as demonstrated in Figure 1.

From equation 1 we see that the projected Einstein ring radius for stellar-mass lenses is much larger than the intra-L2 separation of \euclid\ and \wfirst, so we can expect them to see very similar microlensing events. Nevertheless, the differences between the events may still be observable if an event is highly magnified. For lens-source separations normalized to the Einstein radius $u\lesssim 0.5$, the microlensing magnification scales as ${\sim}u^{-1}$, and its gradient is ${\sim}u^{-2}$. Therefore, even small differences in observer location can result in observable differences in magnification with precise enough photometry. For example, for an event with $u=0.1$, a $0.003$~AU \euclid--\wfirst\ baseline would result in a ${\sim}3$\% difference in magnification that is measurable with photometric precision of ${\sim}1$\% or better that is achievable with both spacecraft for a large number of microlensing events.

Detailed Fisher matrix calculations by Bachelet et al. (2019, in prep.) show that for the brightest of \wfirst's source stars, which will be those most likely to prevent mass measurements via direct lens detection, \euclid\ will be able to detect parallax from the host star's microlensing event for hundreds of events. In these events where the lens is directly detectable, the parallax measurement will deliver significantly improved mass measurements, especially for single lenses. These events will provide invaluable cross-checks on \wfirst's lens-detection masses, and provide an estimate of how many such measurements may be compromised by companion or unrelated stars.

In addition to measuring parallaxes, \euclid\ can confirm the planetary nature of some microlensing events that are vulnerable to confusion with a possible false positive. Binary source stars with extreme flux ratios can cause lightcurve anomalies that resemble low-mass planets~\citep{Gaudi1998bs}. However, the binary source anomalies are usually chromatic, compared to the achromatic planetary deviation. Therefore, high cadence observations with \euclid's VIS instrument (see below) can be used to discriminate between the two scenarios, and confirm \wfirst\ planets.\\



\noindent
{\bf {\it Euclid} is an excellent microlensing observatory}

\euclid\ \citep{Laureijs2011} is a 1.3-m telescope with two instruments (VIS and NISP) fed by a dichroic that can image the same $0.5$~deg$^2$ field of view simultaneously. VIS is a diffraction limited optical imager with a single, wide 0.55--0.90~$\mu$m bandpass and 0.1'' pixels. NISP has 0.3'' pixels, with grisms and imaging filters spanning 0.95--1.85$\mu$m. The spacecraft will follow a halo orbit around L2. NASA contributed the infrared detectors to the \euclid\ mission, in return for US participation in the mission.


\euclid\ maneuvers using cold gas thrusters, which produce no jitter when not firing, but result in a slow slew and settle time of $350$~s to move by one field of view, though it is possible that its actual performance will be better than this design requirement. The long slew and settle time limits the cadence at which more than one field can be observed in time series. For a cadence of 15~minutes, matching \wfirst, it is only possible to observe two fields with an exposure time of 100~s, which would cover only half of \wfirst's ${\sim}2$~deg$^2$ microlensing field. Four fields can be covered at ${\sim}30$~minute cadence, or some combination of two or three fields could be observed with longer exposure times. Four \euclid\ fields could cover ${\sim}90$\% of \wfirst's microlensing fields.

Combining simultaneous 100~s exposures in the VIS and NISP H bands, \euclid\ can achieve a photometric signal to noise ratio that is ${\sim}60$--$85$\% of that achieved by a single 47~s \wfirst\ microlensing observation over the magnitude range $W146=20$--$24$ (AB magnitudes in \wfirst's wide 1-2$\mu$m filter), when accounting for colors, crowding, sky backgrounds and extinction in the field~(Bachelet et al. 2019, in prep.). This magnitude range covers the majority of \wfirst's source magnitudes for events with planet detections~\citep{Penny2019}, and so \euclid\ can provide useful parallax measurements for a large fraction of \wfirst's planets provided that it observes them.

During its main mission \euclid\ will operate within a limited range of solar aspect angles $SAA=87$--$110\degr$ to maintain thermal stability~\citep{GomezAlvarez2018}, which limits the time it can observe the Galactic bulge to two 23-day long seasons per year, or ${\sim}32$\% of \wfirst's 72-day microlensing seasons. However, its possible that the spacecraft could operate over a wider range of angles in an extended mission with less stringent stability requirements. If it was possible to expand these limits to be symmetric about $90\degr$ in an extended mission, then roughly $56$\% or more simultaneous coverage with \wfirst\ would be possible. With current launch schedules and survey strategies, half of \wfirst's microlensing seasons would fall within the \euclid\ main mission (2022--2028), and the other half would fall afterwards.

If every possible opportunity for simultaneous high-cadence observations with \euclid\ were taken in the main mission (69 days total), and if $SAA$ constraints were relaxed for an extended mission (enabling 120 days of simultaneous observations), \euclid\ could provide simultaneous observations during ${\sim}44$\% of \wfirst's microlensing observations. If these observations are only useful for half of \wfirst's planets, likely a conservative assumption, \euclid\ could provide parallaxes for ${\sim}70$ free-floating planets and ${\sim}300$ or more bound planets. High-cadence VIS observations would also discriminate between low-mass planet and high-flux-ratio binary source events.

Taking 69 days out of \euclid's 6 year prime mission (${\sim}3$\% of the mission duration) may not be feasible if \euclid\ is to meet its survey requirements. However, we note that due to \wfirst's later launch date, these observations would occur in the second half of \euclid's mission, when the observing schedule will be less efficient and schedule holes are more likely to exist~\citep{GomezAlvarez2018}. These holes are likely to be smaller than the total 23-day \euclid\ bulge season, but even continuous spans of high cadence observations lasting as short as ${\sim}1$~day would be useful. In a 1 day period, \wfirst\ can expect to detect ${\sim}3$ bound planets and ${\sim}0.75$ free-floating planets, and 24 hours of coverage would capture most or all of the planetary signal.\\





\noindent
{\bf Precursor {\it Euclid} observations of the {\it WFIRST} field}

For one third of \wfirst's planets, detected in the microlensing seasons closest to the mission's midpoint, there will only be a baseline of at most 3.5 years over which to measure the lens separating from the source star. For a typical event, this will result in a ${\sim}0.2$ pixel separation between the lens and source, making mass measurement by lens detection challenging. \euclid\ will launch in 2022, three years earlier than \wfirst\ in 2025. {\bf An early observation using \euclid's VIS channel of the entire \wfirst\ microlensing field, possibly during \euclid's commissioning activities, would double the proper motion baseline over which to measure lens-source separations and masses for one third of \wfirst's planet hosts, and extend the baseline for another third.} Early \euclid\ precursor observations of the entire \wfirst\ microlensing field would also allow a cross check of every \wfirst\ mass measurement using a different telescope, instrument, detector technology, and wavelength. They would also act as an insurance policy hedging against a catastrophic failure of \wfirst\ mid-way through its nominal mission lifetime. Finally, a survey by \euclid\ of the planned \wfirst\ microlensing fields three years before launch would allow an improved optimization of the fields.

\euclid's VIS channel has a very similar pixel scale and PSF size to \wfirst's Wide Field Instrument, and a field of view nearly twice the size. Observations of the $>90$\% of the \wfirst\ microlensing field would require five dithered \euclid\ VIS pointings. Eight pointings would cover essentially the entire \wfirst\ field, with some room for the fields to move with field optimization. An 8-field survey by \euclid\ taking 24 hours would reach $S/N\approx 25$--$50$ in VIS for sources with $W146=23.7$, the median magnitude of \wfirst's planet hosts over most of the \wfirst\ microlensing field.  A 96-hour survey would reach up to $S/N\approx50$--$100$.\\

\noindent
{\bf Summary}

Through the provision of infrared detectors by NASA, the US community has a share in the \euclid\ mission. In this white paper we have outlined the ways in which a small-to-modest investment of \euclid\ observing time during its prime mission, and during a possible extended mission, could significantly enhance the scientific yield of \wfirst's exoplanet microlensing survey. \wfirst\ is a major priority of NASA, and the US community as a whole, as expressed in the 2010 decadal survey~\citep{nwnh} and the 2018 NAS Exoplanet Science Strategy report~\citep{nasess2018}.

\euclid\ observations in support of \wfirst's microlensing survey would provide microlensing parallax and hence mass measurements for a significant fraction of \wfirst's bound and free-floating planets. For the free-floating planets, this is the only way in which their masses can be measured. Measurement of the mass distribution of free-floating and wide-orbit planets, as opposed to just their microlensing timescale distribution, would significantly improve our understanding of the outermost regions of planetary systems, the dynamical evolution of planetary systems, and the frequency of planet ejection. Precursor observations by \euclid\ of the \wfirst\ microlensing field would improve \wfirst's mass measurements for bound planets, and provide important cross-checks against possible sources of systematic errors.

\pagebreak

\bibliographystyle{mn2e}
\bibliography{libraryshort,apj-jour}

\begin{thebibliography}{28}
\expandafter\ifx\csname natexlab\endcsname\relax\def\natexlab#1{#1}\fi

\bibitem[{Barclay {et~al}\mbox{.}(2017)Barclay, Quintana, Raymond, \&
  Penny}]{Barclay2017}
Barclay T., Quintana E.~V., Raymond S.~N., Penny M.~T., 2017, ApJ, 841, 86

\bibitem[{Batygin {et~al}\mbox{.}(2019)Batygin, Adams, Brown, \&
  Becker}]{Batygin2019}
Batygin K., Adams F.~C., Brown M.~E., Becker J.~C., 2019, PhR, in press

\bibitem[{Batygin \& Brown(2016)}]{Batygin2016}
Batygin K., Brown M.~E., 2016, AJ, 151, 22

\bibitem[{Batygin {et~al}\mbox{.}(2012)Batygin, Brown, \& Betts}]{Batygin2012}
Batygin K., Brown M.~E., Betts H., 2012, ApJL, 744, L3

\bibitem[{Bennett {et~al}\mbox{.}(2007)Bennett, Anderson, \&
  Gaudi}]{Bennett2007}
Bennett D.~P., Anderson J., Gaudi B.~S., 2007, ApJ, 660, 781

\bibitem[{Bhattacharya {et~al}\mbox{.}(2018)Bhattacharya, Beaulieu, Bennett,
  Anderson, Koshimoto, Lu, Batista, Blackman, Bond, Fukui, Henderson, Hirao,
  Marquette, Mroz, Ranc, \& Udalski}]{Bhattacharya2018}
Bhattacharya A. {et~al.}, 2018, AJ, 156, 289

\bibitem[{{Committee for a Decadal Survey of Astronomy and
  Astrophysics}(2010)}]{nwnh}
{Committee for a Decadal Survey of Astronomy and Astrophysics}, 2010, {New
  Worlds, New Horizons in Astronomy and Astrophysics}. Tech. rep.

\bibitem[{Gaudi(1998)}]{Gaudi1998bs}
Gaudi B.~S., 1998, ApJ, 506, 533

\bibitem[{Gomes {et~al}\mbox{.}(2005)Gomes, Levison, Tsiganis, \&
  Morbidelli}]{Gomes2005}
Gomes R., Levison H.~F., Tsiganis K., Morbidelli A., 2005, Natur, 435, 466

\bibitem[{G{\'{o}}mez-{\'{A}}lvarez
  {et~al}\mbox{.}(2018)G{\'{o}}mez-{\'{A}}lvarez, Dupac, Buenadicha, Vavrek,
  Hoar, Laureijs, Scaramella, Cuillandre, Dinis, Amiaux, \&
  Tereno}]{GomezAlvarez2018}
G{\'{o}}mez-{\'{A}}lvarez P. {et~al.}, 2018, SPIE, 10707, 1070712

\bibitem[{Gould(2000)}]{Gould2000}
Gould A., 2000, ApJ, 542, 785

\bibitem[{Laureijs {et~al}\mbox{.}(2011)Laureijs, Amiaux, Arduini,
  Augu{\`{e}}res, Brinchmann, Cole, Cropper, Dabin, Duvet, Ealet, Garilli,
  Gondoin, Guzzo, Hoar, Hoekstra, Holmes, Kitching, Maciaszek, Mellier, Pasian,
  Percival, Rhodes, {Saavedra Criado}, Sauvage, Scaramella, Valenziano, Warren,
  Bender, Castander, Cimatti, {Le F{\`{e}}vre}, Kurki-Suonio, Levi, Lilje,
  Meylan, Nichol, Pedersen, Popa, {Rebolo Lopez}, Rix, Rottgering, Zeilinger,
  Grupp, Hudelot, Massey, Meneghetti, Miller, Paltani, Paulin-Henriksson,
  Pires, Saxton, Schrabback, Seidel, Walsh, Aghanim, Amendola, Bartlett,
  Baccigalupi, Beaulieu, Benabed, Cuby, Elbaz, Fosalba, Gavazzi, Helmi, Hook,
  Irwin, Kneib, Kunz, Mannucci, Moscardini, Tao, Teyssier, Weller, Zamorani,
  {Zapatero Osorio}, Boulade, Foumond, {Di Giorgio}, Guttridge, James, Kemp,
  Martignac, Spencer, Walton, Bl{\"{u}}mchen, Bonoli, Bortoletto, Cerna,
  Corcione, Fabron, Jahnke, Ligori, Madrid, Martin, Morgante, Pamplona, Prieto,
  Riva, Toledo, Trifoglio, Zerbi, Abdalla, Douspis, Grenet, Borgani, Bouwens,
  Courbin, Delouis, Dubath, Fontana, Frailis, Grazian, Koppenh{\"{o}}fer,
  Mansutti, Melchior, Mignoli, Mohr, Neissner, Noddle, Poncet, Scodeggio,
  Serrano, Shane, Starck, Surace, Taylor, Verdoes-Kleijn, Vuerli, Williams,
  Zacchei, Altieri, {Escudero Sanz}, Kohley, Oosterbroek, Astier, Bacon,
  Bardelli, Baugh, Bellagamba, Benoist, Bianchi, Biviano, Branchini, Carbone,
  Cardone, Clements, Colombi, Conselice, Cresci, Deacon, Dunlop, Fedeli,
  Fontanot, Franzetti, Giocoli, Garcia-Bellido, Gow, Heavens, Hewett, Heymans,
  Holland, Huang, Ilbert, Joachimi, Jennins, Kerins, Kiessling, Kirk, Kotak,
  Krause, Lahav, van Leeuwen, Lesgourgues, Lombardi, Magliocchetti, Maguire,
  Majerotto, Maoli, Marulli, Maurogordato, McCracken, McLure, Melchiorri,
  Merson, Moresco, Nonino, Norberg, Peacock, Pello, Penny, Pettorino, {Di
  Porto}, Pozzetti, Quercellini, Radovich, Rassat, Roche, Ronayette, Rossetti,
  Sartoris, Schneider, Semboloni, Serjeant, Simpson, Skordis, Smadja, Smartt,
  Spano, Spiro, Sullivan, Tilquin, Trotta, Verde, Wang, Williger, Zhao,
  Zoubian, \& Zucca}]{Laureijs2011}
Laureijs R. {et~al.}, 2011, arXiv:1110.3193

\bibitem[{Levison {et~al}\mbox{.}(2008)Levison, Morbidelli, Vanlaerhoven,
  Gomes, \& Tsiganis}]{Levison2008}
Levison H., Morbidelli A., Vanlaerhoven C., Gomes R., Tsiganis K., 2008, Icar,
  196, 258

\bibitem[{Mann {et~al}\mbox{.}(2010)Mann, Gaidos, \& Gaudi}]{Mann2010}
Mann A.~W., Gaidos E., Gaudi B.~S., 2010, ApJ, 719, 1454

\bibitem[{Morbidelli {et~al}\mbox{.}(2005)Morbidelli, Levison, Tsiganis, \&
  Gomes}]{Morbidelli2005}
Morbidelli A., Levison H.~F., Tsiganis K., Gomes R., 2005, Natur, 435, 462

\bibitem[{Mr{\'{o}}z {et~al}\mbox{.}(2018)Mr{\'{o}}z, Ryu, Skowron, Udalski,
  Gould, Szyma{\'{n}}ski, Soszy{\'{n}}ski, Poleski, Pietrukowicz, Koz{\l}owski,
  Pawlak, Ulaczyk, Albrow, Chung, Jung, Han, Hwang, Shin, Yee, Zhu, Cha, Kim,
  Kim, Kim, Lee, Lee, Lee, Park, Pogge, Pogge, \& Collaboration}]{Mroz2018}
Mr{\'{o}}z P. {et~al.}, 2018, AJ, 155, 121

\bibitem[{Mr{\'{o}}z {et~al}\mbox{.}(2019)Mr{\'{o}}z, Udalski, Bennett, Ryu,
  Sumi, Shvartzvald, Skowron, Poleski, Pietrukowicz, Koz{\l}owski,
  Szyma{\'{n}}ski, Wyrzykowski, Soszy{\'{n}}ski, Ulaczyk, Rybicki, Iwanek,
  Albrow, Chung, Gould, Han, Hwang, Jung, Shin, Yee, Zang, Cha, Kim, Kim, Kim,
  Lee, Lee, Lee, Park, Pogge, Abe, Barry, Bhattacharya, Bond, Donachie, Fukui,
  Hirao, Itow, Kawasaki, Kondo, Koshimoto, Li, Matsubara, Muraki, Miyazaki,
  Nagakane, Ranc, Rattenbury, Suematsu, Sullivan, Suzuki, Tristram, Yonehara,
  Maoz, Kaspi, Friedmann, Group, Maoz, Kaspi, \& Friedmann}]{Mroz2019}
Mr{\'{o}}z P. {et~al.}, 2019, A\&A, 622, A201

\bibitem[{{National Academies of Science, Engineering} \&
  Medicine(2018)}]{nasess2018}
{National Academies of Science, Engineering}, Medicine, 2018, {Exoplanet
  Science Strategy}. National Academies Press, Washington, D.C.

\bibitem[{Nesvorny(2011)}]{Nesvorny2011}
Nesvorny D., 2011, ApJL, 742, L22

\bibitem[{Penny {et~al}\mbox{.}(2019)Penny, Gaudi, Kerins, Rattenbury, Mao,
  Robin, \& {Calchi Novati}}]{Penny2019}
Penny M.~T., Gaudi B.~S., Kerins E., Rattenbury N.~J., Mao S., Robin A.~C.,
  {Calchi Novati} S., 2019, ApJS, 241, 3

\bibitem[{Penny {et~al}\mbox{.}(2013)Penny, Kerins, Rattenbury, Beaulieu,
  Robin, Mao, Batista, {Calchi Novati}, Cassan, Fouque, McDonald, Marquette,
  Tisserand, \& {Zapatero Osorio}}]{Penny2013}
Penny M.~T. {et~al.}, 2013, MNRAS, 434, 2

\bibitem[{Pfyffer {et~al}\mbox{.}(2015)Pfyffer, Alibert, Benz, \&
  Swoboda}]{Pfyffer2015}
Pfyffer S., Alibert Y., Benz W., Swoboda D., 2015, A\&A, 579, A37

\bibitem[{Rasio \& Ford(1996)}]{Rasio1996}
Rasio F.~A., Ford E.~B., 1996, Sci, 274, 954

\bibitem[{Spergel {et~al}\mbox{.}(2015)Spergel, Gehrels, Baltay, Bennett,
  Breckinridge, Donahue, Dressler, Gaudi, Greene, Guyon, Hirata, Kalirai,
  Kasdin, Macintosh, Moos, Perlmutter, Postman, Rauscher, Rhodes, Wang,
  Weinberg, Benford, Hudson, Jeong, Mellier, Traub, Yamada, Capak, Colbert,
  Masters, Penny, Savransky, Stern, Zimmerman, Barry, Bartusek, Carpenter,
  Cheng, Content, Dekens, Demers, Grady, Jackson, Kuan, Kruk, Melton, Nemati,
  Parvin, Poberezhskiy, Peddie, Ruffa, Wallace, Whipple, Wollack, Zhao,
  Spergel, Gehrels, Baltay, Bennett, Breckinridge, Donahue, Dressler, Gaudi,
  Greene, Guyon, Hirata, Kalirai, Kasdin, Macintosh, Moos, Perlmutter, Postman,
  Rauscher, Rhodes, Wang, Weinberg, Benford, Hudson, Jeong, Mellier, Traub,
  Yamada, Capak, Colbert, Masters, Penny, Savransky, Stern, Zimmerman, Barry,
  Bartusek, Carpenter, Cheng, Content, Dekens, Demers, Grady, Jackson, Kuan,
  Kruk, Melton, Nemati, Parvin, Poberezhskiy, Peddie, Ruffa, Wallace, Whipple,
  Wollack, \& Zhao}]{Spergel2015}
Spergel D. {et~al.}, 2015, arXiv, 1503.03757

\bibitem[{Street {et~al}\mbox{.}(2018)Street, Lund, Donachie, Khakpash,
  Golovich, Penny, Bennett, Dawson, Pepper, Rabus, Szkody, Clarkson, {Di
  Stefano}, Rattenbury, Hundertmark, Tsapras, Ridgway, Stassun, Bozza,
  Bhattacharya, {Calchi Novati}, \& Shvartzvald}]{Street2018}
Street R.~A. {et~al.}, 2018, arXiv e-prints, arXiv:1812.04445

\bibitem[{Trujillo \& Sheppard(2014)}]{Trujillo2014}
Trujillo C.~A., Sheppard S.~S., 2014, Natur, 507, 471

\bibitem[{Tsiganis {et~al}\mbox{.}(2005)Tsiganis, Gomes, Morbidelli, \&
  Levison}]{Tsiganis2005}
Tsiganis K., Gomes R., Morbidelli A., Levison H.~F., 2005, Natur, 435, 459

\bibitem[{Weidenschilling \& Marzari(1996)}]{Weidenschilling1996}
Weidenschilling S.~J., Marzari F., 1996, Natur, 384, 619

\end{thebibliography}

\end{document}